# Cascaded Stokes and anti-Stokes laser based on an optical resonator with a self-assembled organic monolayer

ANDRE KOVACH,[1] ARYNN GALLEGOS,[2] JINGHAN HE,[3] HYUNGWOO CHOI,[1] ANDREA M. ARMANI [1,2,3]*

[1]*Mork Family Department of Chemical Engineering and Materials Science, University of Southern California, Los Angeles, California 90089, USA*
[2]*Ming Hsieh Department of Electrical Engineering-Electrophysics, University of Southern California, Los Angeles, California 90089, USA*
[3]*Department of Chemistry, University of Southern California, Los Angeles, CA 90089 USA*
*\*Corresponding author: armani@usc.edu*

**Due to their high circulating intensities, ultra-high quality factor dielectric whispering-gallery mode resonators have enabled the development of low threshold Raman microlasers. Subsequently, other Raman-related phenomena, such as cascaded stimulated Raman scattering (CSRS) and stimulated anti-Stokes Raman scattering (SARS), were observed. While low threshold frequency conversion and generation have clear applications, CSRS and SARS have been limited by the low Raman gain. In this work, the surface of a silica resonator is modified with an organic monolayer, increasing the Raman gain. Up to four orders of CSRS is observed with sub-mW input power, and the SARS efficiency is improved by three orders of magnitude compared to previous studies with hybrid resonators.**

Over the past two decades, significant research efforts have been invested in developing lasers based on stimulated Raman scattering (SRS). [1-4] Because SRS lasers rely on the excitation of vibrational modes in molecules, using the Raman gain intrinsic to the cavity material offers a fundamental advantage over the more conventional laser designs which rely on the excitation of specific electronic transitions in order to emit. [5, 6] As a result, the SRS emission wavelength is not dictated by a narrow electronic transition, but instead by the broader optical transparency window of the cavity material. [7]

If the efficiency of the SRS process is sufficiently high, cascaded stimulated Raman scattering (CSRS) can be observed. [8] In CSRS, the photons generated by one SRS process act as the pump photons for a higher order SRS process, resulting in equally spaced SRS emissions in the frequency domain with decreasing intensities (Figure 1a). [9, 10] While CSRS allows longer wavelength photons to be generated, to reach shorter wavelengths, stimulated anti-Stokes Raman Scattering (SARS) can be used.

SARS is a four-wave mixing process that combines the initial pump photon with the photon generated by SRS to create an upconverted photon. [11] Therefore, the efficiency of the SARS process and the intensity of the SARS emission are directly dependent on the SRS process. As shown in Figure 1a, the frequency difference between the pump photon and both the SRS and SARS photons is fixed and is determined by the vibrational mode of the molecule being excited, also called the Stokes shift.

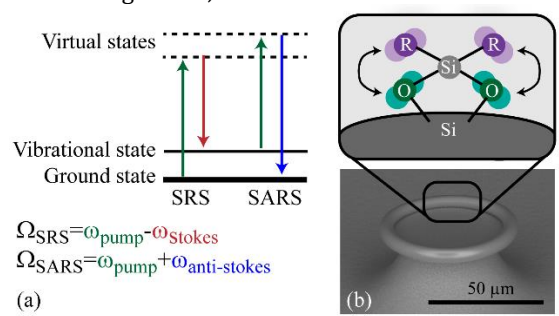

Fig. 1. (a) Energy diagram including both SRS and SARS processes. (b) SEM image of the silica microtoroid resonator with a rendering illustrating molecular vibrations of covalently bound organic molecules on the surface. The "R" groups can represent any general alkyl group covalently bound to oxygen.

Given the ability to simultaneously up- and down-convert pump photons, laser systems based on Raman are of particular interest to researchers in imaging and in spectroscopy. As a result, numerous design strategies based on high efficiency optical cavities and high Raman gain materials have been pursued. One initial effort focused on reducing the lasing threshold and relied on leveraging the ultra-high quality (Q) factors achievable in silica whispering gallery mode resonators. [12] Specifically, in microtoroid optical cavities, the SRS threshold power ($P_{th}$) is directly related to the cavity quality factor. At critical coupling, it can be approximated as [13, 14]

$$P_{th} = \frac{\pi^2 n^2 V_{eff}}{g_R \lambda_P \lambda_R} \left(\frac{1}{Q_o}\right)^2 \frac{(1+K)^3}{K}. \quad (1)$$

where $V_{eff}$ is the effective mode volume, $g_R$ is the Raman gain, K is the normalized coupling parameter, $Q_o$ is the intrinsic quality factor of the resonator, and $\lambda_P$ and $\lambda_R$ are the pump photon wavelength and Raman photon wavelength, respectively. Therefore, higher cavity Q factors directly translate to lower SRS threshold powers, and this process allowed many devices to achieve sub-mW and sub-μW thresholds as well as cascaded SRS lasing, even in materials with low Raman gain coefficients. [15, 16]

However, a low threshold is not sufficient for SARS lasing to be realized. As can be seen in equation (2), the intensity of the anti-Stokes ($I_{SARS}$) is related to the intensity of the SRS ($I_{SRS}$) and pump ($I_{Pump}$) emission lines: [17]

$$I_{SARS} = \left(\frac{4 w_p n_2 t}{\varepsilon c}\right)^2 \left(\frac{\sin(\Delta \omega t / 2)}{\Delta \omega t / 2}\right)^2 I_{Pump}^2 \frac{I_{SRS}}{A_{eff}^2} \quad (2)$$

where $\omega_P$ is the pump frequency, $n_2 \approx 2.2 \times 10^{-20} \ m^2/W$ is the Kerr nonlinear coefficient for silica, c is the speed of light in vacuum, and $A_{eff}$ is the effective mode area. $\Delta \omega = 2\omega_P - \omega_R - \omega_A$ is the frequency detuning, where $\omega_R$ and $\omega_A$ are the frequency of the Stokes and the anti-Stokes modes, respectively, and $t$ is the interaction time between the pump and the Raman modes. Above the Raman threshold, a clamped pump field is observed, and $I_{Pump}$ can be assumed to be a constant which is independent of the coupled power. Therefore, the most straight-forward way to improve the $I_{SARS}$ is to increase the efficiency of the SRS process. However, the lasing efficiencies of many of the initial resonant devices fabricated from a monolithic material system were under 10%. As a result, achieving SARS was extremely challenging at mW input powers without doping the cavity material.

One emerging strategy to overcome these limitations is to design a multi-material device comprised of an integrated optical cavity with a secondary coating. [18] This approach leverages the evanescent field that is intrinsic to whispering gallery mode devices. Initial demonstrations combined silica sol-gel coatings with integrated toroidal cavities, and in more recent work, it was shown that by self-assembling a monolayer of methyl groups on the surface of an optical cavity, the lasing efficiency of a stimulated Raman laser could be increased from ~5% to over 40%. [19] This past work set the stage for the present study.

Using silica toroidal cavities functionalized with self-assembled monolayers of methyl groups, cascaded Raman lasing is demonstrated at 765nm. When compared to non-functionalized silica cavities, the methyl cavities were able to generate a higher number of cascades with less input power. In addition, the efficiency and threshold of cascade formation demonstrated polarization dependence. Lastly, over the range of input powers investigated, only the methyl-functionalized cavities were able to generate SARS due to the higher efficiency of the SRS process.

The initial silica toroidal cavities are fabricated using a standard process comprised of photolithography, buffered oxide and xenon difluoride etching, and $CO_2$ laser reflow. The final major (minor) radius of the devices is 27.5 (3) μm. To functionalize the device surface, a silanization protocol is used in which the device surface is first treated with an oxygen plasma to increase the surface density of hydroxyl groups (-OH). Then, using chemical vapor deposition (CVD), the methyl-containing organic reagent (methyltrichlorosilane, MS) is deposited. The CVD process occurs at room temperature for 10 minutes under Argon purge. During the deposition, MS reacts with the surface-bound hydroxyl groups, forming a monolayer of methyl groups anchored to the device surface via a Si-O-Si bond (Figure 1b). This chemistry process has been shown to be generalizable for the attachment of a range of functional groups to toroidal surfaces. [19-21]

To characterize the device, light from a narrow linewidth laser centered at 765nm (Velocity series, Newport) was evanescently coupled into and out of the cavity using a tapered fiber waveguide (Figure 2). An inline fiber polarization controller was inserted between the laser and the cavity. The waveguide was aligned with the cavity using a three axis nano-positioning stage, and the coupled power was monitored in real-time on an oscilloscope. The output signal was split and sent to an oscilloscope (10%) and an optical spectrum analyzer (OSA) (90%). By tuning the laser, a cavity resonance was identified, and the resonance spectrum was recorded. The intrinsic cavity Q was determined by measuring the linewidth ($\delta\lambda$) over a range of loading conditions and by using a coupled cavity model. [22]

The SRS and SARS spectra were recorded on the OSA. The spectra were measured over a range of input powers and at two different input polarizations to determine the polarization-dependence of the threshold. While convention is to report the threshold power, this metric is dependent on several device variables, including Q and mode area. Therefore, to compare across cavities, the more universal and device-independent metric, circulating intensity at threshold ($I_{circ, thres} = P_{thres}Q/A_{eff}$), is used. The lasing efficiency is the slope efficiency and is determined from the lasing threshold curve.

It is important to note that because all lasing characterization measurements are uni-directional, they set a lower bound on the lasing efficiency and an upper bound on the lasing threshold.

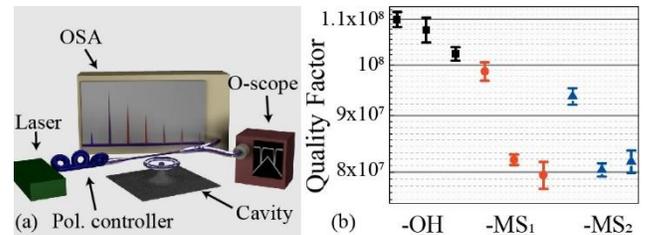

Fig. 2. (a) Setup layout for characterizing SRS and SARS phenomena. (b) Intrinsic quality factors of a series of: hydroxy (-OH) and methyl siloxy functionalized devices at different input polarizations ($MS_1$ and $MS_2$).

As shown in Figure 2(b), the optical Q factors of the cavities ranged between $7.9 \times 10^7$ to $9.8 \times 10^7$. Therefore, the addition of the methyl silane surface coating had minimal impact on the cavity Q. For reference, assuming an input power of 1mW and all other variables as previously defined, this Q range corresponds to a circulating intensity in the cavity between 2.87 $GW/cm^2$ to 3.57 $GW/cm^2$.

Figure 3 presents a representative series of OSA spectra for the two device types studied. The coupled optical power was similar in all measurements, approximately 0.4mW. While all spectra contain cascaded SRS emission lines, the functionalized devices were able to support four cascades with higher intensities than the non-functionalized devices. Similarly, emissions due to SARS is observed in the functionalized devices, but it is absent in the unfunctionalized devices. In addition, the Stokes and anti-Stokes shifts for a specific device were identical, within experimental error. Across devices studied, the values ranged between 12.3 THz to 15.4 THz. These shifts fall within the expected range for the Si-O-Si bond. [23]

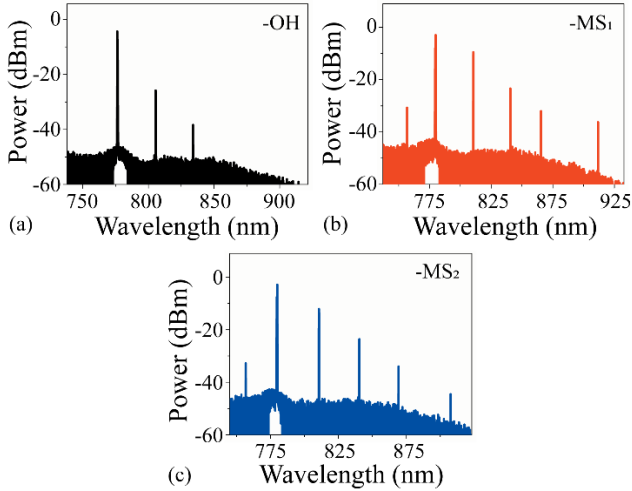

Fig. 3. Optical spectrum of (a) non-functionalized silica device, (b) methyl siloxy functionalized device at polarization state 1, and (c) methyl siloxy functionalized device at polarization state 2. The input power is approximately 0.4mW in all measurements with similar Q values.

To more rigorously quantify the improvement due to the surface chemistry and investigate the role of polarization, threshold curves of the SRS emissions were measured at two different polarizations. These results are shown in Figure 4 (a)-(c). The SRS threshold circulating powers and lasing efficiencies for each device type are determined from the x-intercept and the slopes of these plots. The threshold power is converted into a circulating intensity as previously described, and all results are summarized in Figure 4 (d). The measurements were performed in additional devices with similar results.

Several conclusions can be drawn from Figure 4. The non-functionalized devices are polarization independent, and they have a threshold of 0.298mW ± 0.0103mW and an efficiency of 5% ± 0.036%. This range and behavior are similar to previously reported values. [19] With the addition of the methyl layer, the SRS lasing threshold decreases to 0.148mW and 0.139mW for the two input polarizations investigated. In addition, the efficiency increases to 16% ± 2.1% and 39% ± 3.3%. This represents a maximum decrease in threshold of ~50% and a maximum increase in efficiency of ~8x. Additionally, due to the high efficiency of the CSRS process, the first order emissions in the MS-functionalized devices become nonlinear at approximately 1.5mW of input power. Lastly, the performance becomes polarization dependent. The increase in the SRS lasing efficiency and the appearance of polarization-dependent behavior is directly related to the alignment of the Si-O-Si molecular layer on the surface.

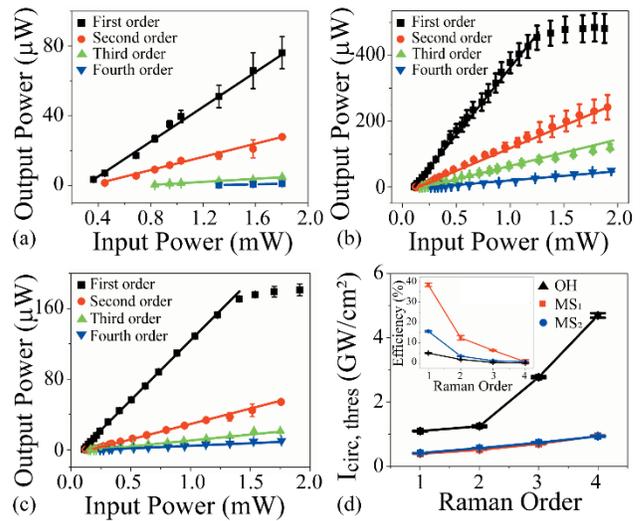

Fig. 4. Threshold behavior of the different devices studied. (a) OH-functionalized device, (b) MS-functionalized device at first polarization state (MS$_1$), and (c) MS-functionalized device at second polarization state (MS$_2$). (d) Circulating intensity at threshold as a function of observed Raman order for all device types. Inset: Efficiency as a function of observed Raman order.

The ability to generate CSRS at low input powers is dependent on the efficiency of the SRS process. Therefore, the combination of increased efficiency and reduced threshold enables four cascades of SRS to be observed with only 0.25mW of input power, as compared to 2mW for the non-functionalized devices, representing roughly an order of magnitude improvement.

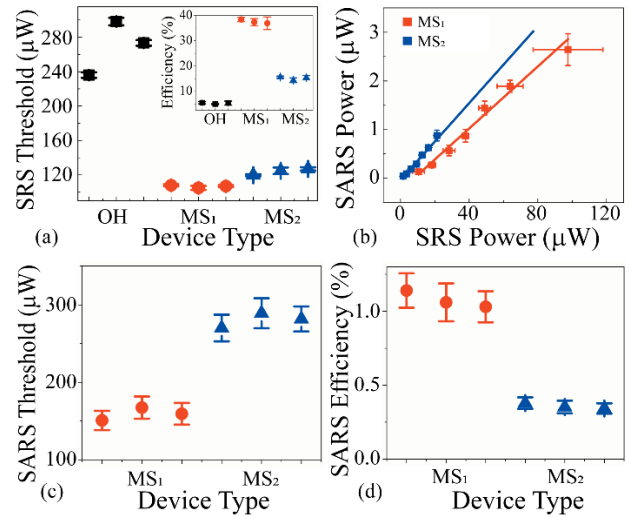

Fig. 5. (a) SRS Threshold for all device types tested. Inset: Efficiency of corresponding SRS (b) SARS power vs SRS power for MS$_1$ and MS$_2$. (c) Threshold and (d) efficiency values of SARS in MS coated devices at both polarizations. SARS was not observed in the non-functionalized devices used in this work.

With input powers as low as ~0.20mW emission lines blue-shifted from the pump generated by SARS were also observed in the methyl siloxy coated devices (Figure 5(a)). The observed linear dependence between the SARS emission intensity and SRS signal (Figure 5(b)) agrees with the previously discussed theory, and this behavior supports that this blue-shifted signal is the result of SARS. In addition, the efficiency of SARS generation reached 1.14% ± 0.48% and the threshold was as low as 0.148mW ± 0.022mW, when the optimum polarization was used (Figure 5(c, d)).

Unfortunately, even with input powers as high as ~2mW, SARS was unable to be observed in the non-functionalized devices. Therefore, direct comparisons between the two device types used in this work cannot be made. In previous work in silica toroidal devices, the highest reported SARS efficiencies are 0.000124% ± 0.000013% with thresholds as low as 1.16mW ± 0.020mW. [17] In related work using multi-material resonator systems, the efficiency and threshold values as high as 0.00151% ± 0.00021% and as low as 0.562mW ± 0.015mW have been achieved. [17] Therefore, in comparison with prior work, the present efficiency results represent a nearly four orders magnitude improvement over non-functionalized devices, and a three orders of magnitude improvement in previous hybrid cavity systems.

The improvement in SARS performance is directly attributable to the improvement in SRS efficiency and SRS threshold. Previous work demonstrated that the Si-O-Si Raman gain was enhanced by forming a monolayer of covalently attached methyl siloxy groups. Further contributions to the SARS enhancement are likely due to a higher number of molecules in the system being able to populate the second virtual state due to the collective vibrational enhancement being localized near the surface and ordered monolayer interface. Smaller contributions could also be from a slight flattening in the cavity dispersion, somewhat relaxing the strict phase-matching conditions for SARS to occur.

In conclusion, by functionalizing the surface of a silica toroidal cavity with a methyl siloxy monolayer, a nearly four orders of magnitude increase in SARS efficiency is demonstrated. This improvement is the result of an increase in SRS efficiency and reduction in SRS threshold. In parallel, CSRS is also demonstrated. In comparison with non-functionalized devices which require 1.3mW of power to generate four order of CSRS, functionalized devices can generate the same number of cascades with only 0.250mW input power. Future work will explore the different contributions to the SARS enhancement as well as the role of the monolayer on the device dispersion. Integrated sources combining SARS and CSRS will be beneficial to many applications, such as in the development of on-chip biosensors and fundamental experiments in cavity-quantum electrodynamics. [24-28]

**Funding.** Army Research Office (ARO) (W911NF1810033) and Office of Naval Research (ONR) (N00014-17-2270).

**Acknowledgment**. We thank Dongyu Chen for assistance with optical mode area calculation.

**Disclosures**. The authors declare no conflicts of interest.